\documentclass[iop]{emulateapj}                             
\usepackage[]{algorithm2e}

\usepackage{amsmath}

\newcommand{\source}{PKS 1939--315}
\newcommand{\mods}{}
\newcommand{\momods}{}
\newcommand{\momomods}{}

\slugcomment{Submitted to ApJ 10th December 2015}

\shorttitle{Mapping Plasma Lenses}
\shortauthors{Tuntsov et al}

\begin{document}

\title{Dynamic spectral mapping of interstellar plasma lenses}

\author{Artem V. Tuntsov$^1$, Mark A. Walker$^2$}
\affil{Manly Astrophysics, 3/22 Cliff Street, Manly 2095, Australia}
\email{1. Artem.Tuntsov@manlyastrophysics.org}
\email{2. Mark.Walker@manlyastrophysics.org}
\author{Leon V.E. Koopmans$^3$}
\affil{Kapteyn Astronomical Institute, University of Groningen, PO Box 800, NL-9700 AV Groningen, The Netherlands}
\email{3. koopmans@astro.rug.nl}
\author{Keith W. Bannister, Jamie Stevens, Simon Johnston}
\affil{CSIRO Astronomy and Space Science, PO Box 76, Epping NSW 1710, Australia}
\author{Cormac Reynolds, Hayley E. Bignall}
\affil{International Centre for Radio Astronomy Research - Curtin University, Perth, Australia}

\begin{abstract}
Compact radio sources sometimes exhibit intervals of large, rapid changes in their flux-density, due to lensing by interstellar plasma crossing the line-of-sight. A novel survey program has made it possible to discover these ``Extreme Scattering Events'' (ESEs) in real time, resulting in a high-quality dynamic spectrum of an ESE observed in \source. Here we present a method for determining the column-density profile of a plasma lens, given only the dynamic radio spectrum of the lensed source, under the assumption that the lens is either axisymmetric or totally anisotropic. Our technique relies on the known, strong frequency dependence of the plasma refractive index in order to determine how points in the dynamic spectrum map to positions on the lens. We apply our method to high-frequency (4.2-10.8~GHz) data from the Australia Telescope Compact Array of the \source\ ESE. The derived electron column-density profiles are very similar for the two geometries we consider, and both yield a good visual match to the data. However, the fit residuals are substantially above the noise level, and deficiencies are evident when we compare the predictions of our model to lower-frequency (1.6-3.1~GHz) data on the same ESE, thus motivating future development of more sophisticated inversion techniques.

\end{abstract}


\section{Introduction}

Extreme Scattering Events (ESEs) were discovered almost thirty years ago \citep{Fiedler1987}, yet their cause remains obscure. That is in part because it is difficult to construct acceptable physical models for the phenomenon. Refraction by ionised interstellar gas appears to be the mechanism responsible for the variations \citep{Fiedler1987, rbc1987}, but the inferred properties of the plasma lenses are problematic. In particular the very high gas pressures, confined in very small regions, are challenging to interpret. And the challenge is heightened by the fact that the lenses are commonplace in the Galaxy, with $\sim10^4\,{\rm pc^{-3}}$ local to the Sun. 

Despite their large numbers, the small size of the lenses (a few AU) means that the event rate is actually quite low; \cite{Fiedler1994} estimated roughly one in 200 sources undergoing an ESE at any instant, but using more restrictive criteria other authors have suggested a fraction as small as one in 2$,$000 (e.g. \citealt{karastergiouwalker2011}). That is the other reason why ESEs are still not understood: we have had very few examples to study. However, that problem is now being addressed with a new survey program which recognises events in progress, and is thus able to follow-up each event with intensive multi-wavelength monitoring \citep{bigpaper}. As a result of that program, we now have a high-quality dynamic radio spectrum, from the Australia Telescope Compact Array (ATCA), for an ESE towards the source \source. Our data span a factor $\sim7$ in wavelength, corresponding to a factor $\sim50$ in the focal length of the lens, and thus provide strong constraints on viable lens models. This paper details our analysis of that dynamic spectrum.

Previous analyses of ESE light-curves have proceeded by fitting to models based on analytic lens profiles, notably the Gaussian form (e.g. \citealt{cfl1998}). Here, however, we attempt to invert the dynamic spectrum of the lensed source and thus discover the column-density profile of the plasma lens without prejudice. To do so we restrict attention to two possible geometries: axisymmetric lenses, and extremely anisotropic lenses (i.e. lenses which refract in only one direction). Most of the physical scenarios which have been proposed for ESEs can be approximated with models which fall into one of these two categories \citep{henriksenwidrow1995,walkerwardle1998,walker2007,penking2012}. And for these geometries the inversion is strongly over-constrained because the profile we seek is one-dimensional, whereas the data are two-dimensional.  

This paper is organised as follows. In section~\ref{section:lensing} we recap the basics of lensing theory, using geometric optics and the thin lens approximation, and apply the theory to the case of a plasma lens. Section~\ref{section:methods} then presents our approach to reconstructing the plasma density in the lens, using the well-defined scalings of the plasma refraction angle and image magnification with wavelength.  Section~\ref{section:application} applies these ideas to our data on \source\ and presents our main results, and section~\ref{section:discussion} discusses their interpretation.

\section{General lens model}
\label{section:lensing}
In the geometric optics approximation (see {\it e.g.}, \citealt{sef1992} where most of the material in this section can be found), the optical effects of a plasma lens can be encapsulated in the ``lens equation''
\begin{equation}\label{dangle}
\boldsymbol{\beta}=\boldsymbol{\theta}-{{D_{ds}}\over{D_s}} \boldsymbol{\alpha},
\end{equation}
which relates the angular position of the source, $\boldsymbol{\beta}$, to the angular position of its image, $\boldsymbol{\theta}$, via the ``deflection angle,'' $\boldsymbol{\alpha}$, and the distances from lens-to-source ($D_{ds}$), and observer-to-source ($D_s$). The deflection angle $\boldsymbol{\alpha}$ is actually the negative of the refraction angle introduced by the lens, with the sign chosen for consistency with the majority of the gravitational lensing literature. In general there may be more than one image of the source, corresponding to the case where there are multiple solutions to equation~(\ref{dangle}) for a given value of $\boldsymbol{\beta}$.

For a thin lens, the refraction can be characterised in terms of a dimensionless scalar quantity, $\Psi$, referred to as the ``lens potential'', such that
\begin{equation}\label{Psidef}
\boldsymbol{\alpha}={{D_{s}}\over{D_{ds}}}\boldsymbol{\nabla}\Psi,
\end{equation}
where $\boldsymbol{\nabla}$ is the gradient with respect to angular position $\boldsymbol\theta$ in the lens plane \citep{schneider1984}. Our lens equation then takes on the simple form
\begin{equation}
\boldsymbol{\beta}=\boldsymbol{\theta}-\boldsymbol{\nabla}\Psi, \label{lenseq}
\end{equation}
and $\Psi$ provides a complete description of the lensing behaviour. We will see, below, that in the case of plasma lensing the potential function is proportional to the electron column-density profile of the lens, $N_e(\boldsymbol{\theta})$.

For sufficiently small sources, and we assume a point source for the rest of this paper, the flux of an image is determined by the magnification of the lens at the location $\boldsymbol{\theta}$, because intensity is conserved along rays if there is no absorption or gain introduced by the lens. Here the term ``magnification'' means the ratio of  solid-angular size of the image to the solid-angular size of the source. Magnification is thus calculated from the differential properties of the lens mapping, i.e. the relationship of $\Delta\boldsymbol{\theta}$ to $\Delta\boldsymbol{\beta}$, where the latter is the locus that describes the perimeter of the source. In general the image is distorted, with different amounts of stretching in different directions, and one evaluates the image magnification $\mu$ from the Jacobian matrix, $\boldsymbol{\rm A}$ \citep{schneider1985}:
\begin{equation}\label{deformatrix}
\boldsymbol{\rm A}\equiv{{\partial\boldsymbol{\beta}}\over{\partial\boldsymbol{\theta}}}=\boldsymbol{\rm 1}-\partial_{\theta_i}\partial_{\theta_j}\Psi,\hspace{.3cm} i,j=\overline{1,2}
\end{equation}
via
\begin{equation}
\mu=\frac{1}{|{\rm Det}(\boldsymbol{\rm A})|}=\frac1{\left|(1-\kappa)^2-\gamma^2\right|}, \label{mag}
\end{equation}
where the convergence, $\kappa$, and shear, $\gamma$, are invariants of $\boldsymbol{\rm A}$,
\begin{equation}\label{kappagammadef}
\kappa\equiv\frac12\left(\Psi_{11}+\Psi_{22}\right),\hspace{.1cm} \gamma^2\equiv\frac14\left(\Psi_{11}-\Psi_{22}\right)^2+\Psi_{12}\Psi_{21},
\end{equation}
familiar from optics. The former describes the isotropic deformation of a small light beam due to the lens, whereas the latter characterises stretching of the beam along one axis relative to the other.
Equations~(\ref{dangle}-\ref{kappagammadef}) are quite general and can be used to describe various types of lensing in the geometric optics approximation. 

\subsection{Potential function of a plasma lens}
The additional phase, $\Phi$, impressed by the lens on a wave which has propagated through it, is given by
\begin{equation}\label{Phidef}
\Phi={{2\pi}\over\lambda}\int\!{\rm d}z\,({\rm n}-1),
\end{equation}
where ${\rm n}$ is the refractive index of the lens, and we have assumed propagation in the $\hat{\boldsymbol{z}}$ direction. Here we are using the thin lens approximation, in which one neglects the effect of deflection of the wave during passage through the lens, and $\Phi=\Phi(x,y)$, independent of the angle of incidence.

As noted above, the angle $\boldsymbol{\alpha}$ is the negative of the refraction angle introduced by the lens. By definition the wavevector is normal to the wavefront, i.e. normal to surfaces of constant phase, and for small deflections the refraction angle is thus given by the ratio of the transverse phase gradient to the longitudinal one, so
\begin{equation}\label{plasmadangle}
\boldsymbol{\alpha}=-{{\lambda}\over{2\pi}} \boldsymbol{\nabla}_{\!r}\Phi.
\end{equation}
Here the operator  $\boldsymbol{\nabla}_{\!r}$ is a two-dimensional vector quantity, and the subscript $r$ emphasises that this gradient is taken with respect to the position $\boldsymbol{r}=D_d\boldsymbol{\theta}$, where $D_d$ is the distance of the lens from the observer. We can relate this operator to the angular gradient via $D_d\boldsymbol{\nabla}_{\!r}=\boldsymbol{\nabla}$.

For an electromagnetic wave of angular frequency $\omega$, the refractive index of a cold plasma is ({\it e.g.}, \citealt{stix1992})
\begin{equation}\label{plasman}
{\rm n}=1-{{\omega_p^2}\over{2\omega^2}},
\end{equation}
where the plasma frequency, $\omega_p$, is assumed to be such that $\omega_p^2\ll\omega^2$. The plasma frequency is given in terms of the electron density, $n_e$, by
\begin{equation}\label{omegapdef}
\omega_p^2={{n_e e^2}\over{m_e \epsilon_o}}.
\end{equation}
Here $e$ and $m_e$ are the charge and mass of the electron, respectively, and $\epsilon_o$ is the permittivity of free-space.

Using equations~(\ref{Phidef}-\ref{plasman}) we thus obtain 
\begin{equation}\label{PhiNe}
\Phi=-r_e \lambda N_e,
\end{equation}
where $r_e$ is the classical radius of the electron. Combining this result with equation~(\ref{Phidef}) then yields the potential function
\begin{equation}\label{psidef}
\Psi=\left({{\lambda}\over{\lambda_o}}\right)^2 \psi,
\end{equation}
where
\begin{equation}\label{psiNe}
\psi\equiv{{1}\over{2\pi}}\,{{D_{ds}}\over{D_dD_s}}\,r_e \lambda_o^2\,N_e,
\end{equation}
and $\lambda_o$ is an arbitrarily chosen reference wavelength.

Note that the deflection angle $\boldsymbol\alpha$ and its derivatives are all proportional to $\lambda^2$:
\begin{equation}\label{agkscale}
\boldsymbol\alpha=\left(\frac{\lambda}{\lambda_0}\right)^2\boldsymbol\alpha_0,\hspace{.2cm} \kappa=\left(\frac{\lambda}{\lambda_0}\right)^2\kappa_0, \hspace{.2cm} \gamma=\left(\frac{\lambda}{\lambda_0}\right)^2\gamma_0.
\end{equation}
Thus, a definite scaling should hold for the inverse magnification\footnote{The minus sign is for images of negative parity.} of an image \emph{measured at a fixed position $\boldsymbol\theta$}:
\begin{equation}\label{muscale}
\pm\frac1\mu=1-2\left(\frac{\lambda}{\lambda_0}\right)^2\kappa_0+\left(\frac{\lambda}{\lambda_0}\right)^4\left(\kappa_0^2-\gamma_0^2\right).
\end{equation}
We now discuss the possibility of such measurements.

\section{Interpreting data}
\label{section:methods}

\subsection{Characteristics}
In the limit where $\lambda\rightarrow0$, the lens mapping of equation~(\ref{lenseq}) reduces to the identity $\boldsymbol{\beta}=\boldsymbol{\theta}$, so there is only one image and the magnification of that image is $\mu=1$. As $\lambda$ increases there is still only one image,  formed close to $\boldsymbol{\theta}=\boldsymbol{\beta}$, but the deflection angle becomes greater, and the magnification deviates further from unity. The single-image/high-frequency regime is convenient for two reasons. First, if the lens is so weak that there is only one image present then there is no ambiguity of interpretation: the current source position corresponds to a unique image position, and the measured flux of the source tells us about the magnification -- and via equation~(\ref{deformatrix}) the curvatures of the lens potential -- at that point. Furthermore the image has positive parity, i.e. ${\rm Det}(\boldsymbol{\rm A})>0$ \citep{burke1981}, and it is not necessary to take the absolute value in equation (\ref{mag}). Secondly, if the image magnification is not far from unity then the influence of source structure on the observed fluxes is less likely to be important, and a point-source model can be used in the first instance. 

In the absence of imaging data -- i.e. data which reveal the angular structure of the lensed source -- the argument~$\boldsymbol\theta$ of the potential $\psi(\boldsymbol{\theta})$ is not observationally accessible and needs to be reconstructed along with the potential function from equations~(\ref{lenseq}, \ref{mag}). It is therefore convenient to consider $\boldsymbol{\theta}(\boldsymbol{\beta},\lambda)$ itself as the primary dependent variable when reconstructing the lens. The deflection angle can then be obtained by interpolating between the $(\boldsymbol\theta-\boldsymbol\beta, \boldsymbol\theta)$ pairs and the locus of $\boldsymbol\theta(\boldsymbol\beta_i, \lambda_j)$ clearly defines the portion of the lens plane constrained by the available data. The potential function and electron column density are subsequently found by integrating the deflection angle. Relegating $\boldsymbol\theta$ to the status of a dependent variable has the considerable practical advantage of bypassing the computationally costly solution of the lens equation. It can also provide an insight into the qualitative structure of the data, and a simple visualisation of the predictive capabilities of the model.

The latter is made possible by identifying the locus of points in the data that correspond to the same position in the lens plane. From the lens equation~(\ref{lenseq}) with the deflection angle~(\ref{agkscale}), all $\boldsymbol\beta$ along the line
\begin{equation}\label{lenscharacter}
\boldsymbol\beta=\boldsymbol\theta-\left(\frac{\lambda}{\lambda_0}\right)^2\boldsymbol\alpha_0(\boldsymbol\theta)
\end{equation}
in the domain of our independent variable $\boldsymbol\beta$ project to the same dependent variable $\boldsymbol\theta$ in the image plane for all $\lambda$. In order to compute $\boldsymbol\alpha$, we need to know $\boldsymbol\theta$ at a single point $(\boldsymbol\beta_1, \lambda_1)$ on the line $\boldsymbol\theta(\boldsymbol\beta, \lambda)=\boldsymbol\theta_1$. Substituting the corresponding value of $\boldsymbol\alpha_0$ into equation~(\ref{lenscharacter}), we obtain
\begin{equation}\label{character}
\boldsymbol\beta(\lambda; \,\boldsymbol\beta_1, \lambda_1, \boldsymbol\theta_1)=\boldsymbol\theta_1-\left(\frac\lambda{\lambda_1}\right)^2\left(\boldsymbol\theta_1-\boldsymbol\beta_1\right).
\end{equation}
Thus, knowing the mapping  at one particular position and wavelength determines the solution along the entire line~(\ref{character}) in its domain, and $\boldsymbol\theta[\boldsymbol\beta(\lambda;\, \boldsymbol\beta_1, \lambda_1, \boldsymbol\theta_1), \lambda]\equiv\boldsymbol\theta_1$ by construction. By analogy with the theory of partial differential equations this line can be referred to as a ``characteristic''.  {\mods Later, in Section~\ref{section:application} (Figure~\ref{figure:models}), we will see examples of systems of characteristic lines appropriate to particular lens models.} 

The reference wavelengths in equations~(\ref{psidef}-\ref{lenscharacter}, \ref{character}) need not be equal and it is convenient to use $\lambda_0=1$, which we assume in the following.

\subsection{One-dimensional problems}
\label{subsection:onedproblems}

Two-dimensional (epoch and wavelength) data of the dynamic spectrum over-constrain models of electron density that depend on only one spatial coordinate, such as the two cases we consider -- extremely anisotropic and axially symmetric lens models. Characteristics are useful in defining which multiple data points constrain the same location in the lens.

\subsubsection{Extremely anisotropic case}
\label{subsubsection:anisotropic}

When the electron density depends on only one of the two Cartesian coordinates in the lens plane, $N_e(\boldsymbol\theta)=N_e(\theta_x)$, the deflection angle is always parallel to it and the source position is unaffected in the other direction:
\begin{equation}\label{lllens}
\boldsymbol\theta(\boldsymbol\beta, \lambda):\hspace{4mm} \theta_x(\beta_x, \lambda)=\beta_x+\lambda^2 \partial_{\theta_x}\psi, \hspace{3mm}\theta_y=\beta_y;
\end{equation}
we will only consider the $x$ component, dropping the subscript. Using equation~(\ref{lenseq}), the image magnification is
\begin{equation}\label{llmag}
\mu={\partial_\beta\theta}=\frac1{1-\lambda^2\partial_\theta\alpha_0(\theta)}
\end{equation}
and therefore, on a characteristic, where the argument of the derivative in the denominator is constant, the derivatives, which are measurable as $\mu$, should give the same result when rescaled to the reference wavelength:
\begin{equation}\label{1dmuscale}
\partial_\beta\theta_0=\frac{\partial_\beta\theta}{1/\lambda^2+(1-1/\lambda^2)\partial_\beta\theta}.
\end{equation}
Comparing equation~(\ref{1dmuscale}) to equation~(\ref{muscale}) we also establish the relationship between the optical scalars in the extremely anisotropic case:
\begin{equation}\label{llinv}
\kappa_0=\frac{\partial_\theta\alpha_0}2=\frac{1-1/\partial_\beta\theta}{2\lambda^2},\hspace{.5cm} \gamma_0^2=\kappa_0^2,
\end{equation}
which can be used as a consistency check if $\kappa_0, \gamma_0$ can be obtained independently -- {\it e.g.}, by fitting a biquadratic polynomial~(\ref{muscale}) to the magnifications along the characteristic defined by $\alpha_0$.

\subsubsection{Axially symmetric case}
\label{subsubsection:axisymmetric}

In complete axial symmetry, $N_e(\boldsymbol\theta)=N_e(\theta)$, the deflection is always in the radial direction:
\begin{equation}\label{oolens}
\boldsymbol\theta(\boldsymbol\beta, \lambda):\hspace{4mm} \theta(\beta, \lambda)=\beta+\lambda^2 \partial_{\theta}\psi, \hspace{5mm}\boldsymbol\theta\parallel{\boldsymbol\beta}.
\end{equation}
The relation  between the observable $\mu$ and $\partial_\beta\theta$ is slightly different as one has to account for the curvature of the polar coordinate system when computing magnification:
\begin{equation}\label{oomag}
\mu=\frac\theta\beta\partial_\beta\theta=\frac1{1-\lambda^2\alpha_0/\theta}\frac1{1-\lambda^2\partial_\theta\alpha_0}.
\end{equation}
But as the lens equation~(\ref{oolens}) is formally equivalent to equation~(\ref{lllens}), the scaling~(\ref{1dmuscale}) remains valid along the characteristic in the axisymmetric case. A slight complication of the $\mu$ to $\partial_\beta\theta$ conversion by the $\theta/\beta$ factor presents no difficulty in practice because, in order to consider variation of the derivative along a characteristic, the characteristic's $\theta$ needs to be specified in the first place. For the axisymmetric case, the optical scalars are given by
\begin{eqnarray}\label{ooinv}
\kappa_0=\frac12\left(\frac{\alpha_0}\theta+\partial_\theta\alpha_0\right)=\frac1{\lambda^2}-\frac{\beta/\theta+1/\partial_\beta\theta}{2\lambda^2},\hspace{.3cm}\\\nonumber
\gamma_0^2=\left[\frac12\left(\frac{\alpha_0}\theta-\partial_\theta\alpha_0\right)\right]^2=\left(\frac{1/\partial_\beta\theta-\beta/\theta}{2\lambda^2}\right)^2.
\end{eqnarray}

\subsubsection{Formalisation of one-dimensional problems}
\label{subsubsection:overconstrainedness}
By definition of a characteristic line, $\theta$ is constant along the line. 
We therefore require that the right-hand side of equation~(\ref{1dmuscale}) -- with $\partial_\beta\theta$ computed as $\mu$ or $\mu\beta/\theta$ in the extremely anisotropic and axisymmetric cases, respectively -- is constant on characteristics:
\begin{equation}\label{constalong}
\partial_\beta\theta_0=\mathrm{const~~on~~}\beta(\lambda)=\theta-\left(\frac{\lambda}{\lambda_1}\right)^2\left(\theta-\beta_1\right).
\end{equation}
This relation expresses the over-constrainedness of the model by the data in the one-dimensional case.

\subsection{Charting the characteristic set}
\label{subsection:reconstruction}

The actual characteristic passing through a given point is not known in advance and one can envisage a number of approaches to correctly drawing the family of these lines through the data. The latter task is equivalent to the lens reconstruction problem since each characteristic corresponds to a measurement $\theta(\beta)$ and can be converted into $\alpha_0(\theta)$ by interpolating on $\theta, (\theta-\beta)/\lambda^2$ pairs. There are two constraints to guide the solution process: equation (\ref{constalong})  and equations~(\ref{llmag}, \ref{oomag}); the former expressing the expected behaviour of the data along the characteristics and the latter across them. The relative weights given to the two in a particular combination define the various ways in which the  solution could be attempted.

{\mods
Condition~(\ref{constalong}) can be explicitly resolved with respect to $\theta$ to define the characteristic suggested by the data in the neighbourhood of a given point:
\begin{equation}\label{localcharacter}
\hat\theta(\beta, \lambda)=\beta-\frac{\partial_\beta\theta+\lambda^2\partial_{\lambda^2}\,\partial_\beta\theta}{\partial_\beta\,\partial_\beta\theta},
\end{equation}
but such $\hat\theta$ would not generally be constant along characteristics, rendering this relation of little practical use. The reason is that it implies a consistency condition on the data themselves which, even if met by the underlying signal, will be violated by the random and systematic errors which are inevitable in the measurement process.

A milder version of this method would be to select, among all the possible lines~(\ref{character}) passing through a given point $(\beta,\lambda)$, the one on which the gradient~(\ref{1dmuscale}) is \emph{least variable}. However,  this approach makes no attempt to satisfy equations~(\ref{llmag}, \ref{oomag}) and the resulting $\theta(\beta)$ does not have to be smooth, continuous or even self-consistent. We have found that both of these ``morphological methods'' performed poorly on our data.

The approach that has so far proven most useful, and the one which we use to analyse our data in Section~\ref{section:application}, is as follows.  We first select a pivot point, $(\beta_1, \lambda_1)$, through which we would like to draw a characteristic. With an assumed value of the corresponding position in the lens plane, $\theta_1$, the trajectory of the characteristic in the data domain, i.e. $\beta(\lambda)$, is then fixed by equation~(\ref{character}). Data points which lie on or near this trajectory give us information on the magnification along the characteristic. Scaling to the reference wavelength, we then use these magnifications to  form the average $\langle\partial_\beta\theta_0\rangle_1$ as our estimate of the gradient of the lens mapping at the position $\theta_1$. We then select a new pivot point, $(\beta_2, \lambda_2)$, lying close to, but not on the first characteristic. The corresponding position in the lens plane,  $\theta_2$, is given by
\begin{equation}\label{continuity}
\theta_2=\theta_1+\langle\partial_\beta\theta\rangle_1(\lambda_2)\Delta\beta,
\end{equation}
where $\langle\partial_\beta\theta\rangle_1(\lambda_2)$ is the gradient of the first characteristic evaluated at $\lambda_2$, and $\Delta\beta=\beta_2-\beta(\lambda_2; \beta_1, \lambda_1, \theta_1)$ is the distance along the $\beta$ axis between the new pivot point and the initial characteristic. Knowing $\theta_2$ fixes the entire trajectory $\beta(\lambda)$ of the new characteristic, and so we can use our data to form an estimate of the gradient, $\langle\partial_\beta\theta_0\rangle_2$, for the new characteristic. This process is then iterated, allowing us to cover the data domain with an entire family of characteristics. Once we fix the set of pivot points, the foregoing procedure furnishes us with a unique lens model for each assumed value of $\theta_1$. 

As we have no prior knowledge of the appropriate value of $\theta_1$, we must choose a suitable figure by determining which of the resulting lens models we prefer. This is readily achieved, e.g. by introducing a figure of demerit that measures the root-mean-square difference between the model predictions and the data. The procedures just described are summarised in pseudocode as Algorithm~\ref{algorithm:trials}.

Because the characteristic lines are never parallel to the $\beta$ axis, it is convenient to choose the same wavelength for all pivot points. We chose all pivots to be at 8.4~GHz. And because deflection angles are small at high frequencies, and our data are not uniformly sampled in time, we placed characteristics in one-to-one correspondence with the sampled epochs, $\{\beta_k\}$. Over much of the 4~cm observing band, this choice gives us a close correspondence between the locations of the data samples and the locations of the characteristics.

\begin{algorithm}
 \KwData{Magnification $\mu_{ij}=\mu(\beta_i, \lambda_j)$}
 \KwResult{Deflection curve $\alpha_0(\theta)$}
Construct interpolating function $\overline\mu(\beta, \lambda)$\;
Select a set of characteristic pivot points $\{(\beta_k,\lambda_k)\},\,k=1..K$\;
Select a set of initial values $\{\theta_1^l\},\,l=1..L$ to try\;
\For{$l=1..L$}{
	Set first characteristic parameter $\theta_1=\theta_1^l$\;
	Compute average scaled derivative $\langle\partial_\beta\theta_0\rangle_1$ from \\\hspace{.5cm}magnification data interpolated onto the \\\hspace{.5cm}characteristic $\beta(\lambda)=\theta_1-(\lambda/\lambda_1)^2(\theta_1-\beta_1)$\;
	\For{k=2..K}{
		Compute distance $\Delta\beta$ to next characteristic\;
		Rescale $\langle\partial_\beta\theta_0\rangle_{k-1}$ to $\lambda=\lambda_k$\;
		Compute parameter $\theta_k=\theta_{k-1}+\langle\partial_\beta\theta\rangle_{k-1}(\lambda_k)\Delta\beta$\;
		Compute average scaled derivative $\langle\partial_\beta\theta_0\rangle_k$ along \\\hspace{.5cm}characteristic $\beta(\lambda)=\theta_k-(\lambda/\lambda_k)^2(\theta_k-\beta_k)$\;
	}
	Compute model $\hat\mu_k[\beta_k(\lambda_n),\lambda_n]$ along all characteristics\;
	Interpolate $\hat\mu_{kn}$ onto data sampling points $(\beta_i, \lambda_j)$\;
	Compute demerit $d(\mu_{ij}, \hat\mu_{ij})$\;
	Store demerit, characteristic parameter set $(d,\{\theta_k\})_l$\; 
}
Identify initial value $\theta_1^l$ that results in optimal demerit\;
$\alpha_0(\theta)$ interpolates optimal $[\theta_k, (\theta_k-\beta_k)/\lambda_k^2]$, $k=1..K$.
\smallskip\smallskip
\caption{Averaging along trial characteristic set. The derivative $\partial_\beta\theta$ is obtained from the magnification using equation~(\ref{llmag}) or equation~(\ref{oomag}), for the extremely anisotropic or axisymmetric geometry, respectively, and scales with wavelength according to equation~(\ref{1dmuscale}).}
\label{algorithm:trials}
\end{algorithm}

}

A clear advantage of this method is that there is just a single variable $\theta_1$ with respect to which the model is to be optimised, and yet it takes account of both equations~(\ref{constalong}) and~(\ref{llmag}, \ref{oomag}), even if only in an average sense for the latter. It also performs well in practice, producing reasonable models and doing so stably with respect to variations in the spectral and kinematic models or readjustment of weighting schemes. A serious drawback, however, is that errors in determining the properties of one characteristic propagate into all subsequent characteristics in the inner loop of Algorithm 1. Of particular concern with the real data, which inevitably contain gaps in the coverage, is the possibility of missing episodes of strong (de-)magnification, which strongly deforms the paths on which the derivatives are averaged.

A simple way to mitigate problems associated with gaps in the coverage is to break the dataset into a few separate pieces at suspect positions -- particularly in the longer gaps in the data coverage, where a missed high-maginfication event might lurk. The algorithm is then run on each of the pieces independently. If the results align well, it is unlikely that an important event has been missed and the data can be restitched together for a more reliable reconstruction. If the algorithm returns considerable offset between the end of one reconstructed piece and the beginning of the next, it is a good indication an important event in the light curve has been missed. 

\section{Application to ATCA data on PKS 1939--315}
\label{section:application}

\subsection{Kinematic and spectral model}
\label{subsection:spectrumkinematics}

The algorithm described above deals with the angular position and magnification of the source whereas the available data come in the form of a dynamic spectrum $F(t, \lambda)$. To convert the latter to the former, models of the instrinsic spectrum of the source, $F_0(\lambda)$, and of its motion relative to the Earth-lens system, $\boldsymbol\beta(t)$, have to be assumed. We aimed to keep both models as simple as practical, restricting ourselves to a power-law spectrum and rectilinear motion:
\begin{equation}\label{spectrum}
F_0(\lambda)=F_0\left(\frac{\lambda_0}\lambda\right)^{p},
\end{equation}
and
\begin{equation}\label{kinematics}
\boldsymbol{\beta}(t)=\boldsymbol\beta_0+\frac{\boldsymbol{v}_f}{D_d}(t-t_0).
\end{equation}
Here the effective transverse velocity, $\boldsymbol{v}_f$, is given in terms of the source, lens and Earth velocities ($\boldsymbol{v}_s$, $\boldsymbol{v}_d$, and $\boldsymbol{v}_\oplus$, respectively) by
\begin{equation}\label{veff}
\boldsymbol{v}_{f}={{D_{d}}\over{D_s}} \boldsymbol{v}_s -  \boldsymbol{v}_d +{{D_{ds}}\over{D_s}} \boldsymbol{v}_\oplus 
\end{equation}
\citep{cordesrickett1998}. In the case of a quasar seen through a plasma lens in our own Galaxy, $D_s\simeq D_{ds} \gg D_d$, so $\boldsymbol{v}_{f}\simeq -  \boldsymbol{v}_d + \boldsymbol{v}_\oplus$. It is convenient to choose the epoch $t_0$ so that $\beta_0$ is the impact parameter in the axisymmetric case:  ${\boldsymbol{\beta}}_0\cdot{\boldsymbol{v}}_f=0$. 

In the extremely anisotropic case, only the $x$ component of equation~(\ref{kinematics}) is relevant and the photometry is insensitive to changes in $\boldsymbol\beta_0$ or $t_0$. Further, $v_f/D_d$ only affects the overall scale of the angular coordinates $\beta, \theta$ by setting the time-to-angular-position conversion factor. That too is irrelevant for the photometry, which is determined by the dimensionless $\partial\theta/\partial\beta$ derivatives. As a result, only the normalisation, $F_0$, and slope, $p$, of the intrinsic spectrum need to be specified, and those choices can be optimised, while the angular positions are effectively measured in time units -- {\it i.e.}, $v_f/D_d\equiv1$.

In the case of axial symmetry the observables depend on the magnitude of $\boldsymbol\beta$,
\begin{equation}\label{axialkinematics}
\beta(t)=\sqrt{\beta_0^2+(v_f/D_d)^2\left(t-t_0\right)^2}.
\end{equation}
and it is now only the overall angular scale of the problem, $v_f/D_d$ that remains free. Thus, the parameter inventory of this case includes both the spectral and kinematic portions. Optimisation in this four-dimensional space is slow, so we only optimised with respect to the impact parameter and epoch, $\beta_0, t_0$, adopting the best-fit spectral model of the extremely anisotropic case. The impact parameter and epoch of closest approach, as determined by our optimisation, are given in Table~\ref{table:parameters} along with the parameters of the spectral model.

\begin{table}
\caption{Best fit spectral and kinematic parameters for PKS 1939--315.}
\label{table:parameters}
\centering
\begin{tabular}{c|c|c|c|c}
$F_0(10\,\mathrm{GHz})$, mJy & $p$ & $\beta_0$, days & $t_0$, MJD& $f$, \% \cr
\hline
 $176\pm3$ & $-0.22\pm0.02$ & $\lesssim5$ & $56790^{+5}_{-10}$ &$5\pm1$ \cr
\end{tabular}
\end{table}

\subsection{Implementation details}
\label{subsection:technicalities}

\begin{figure}
\includegraphics[height=70mm]{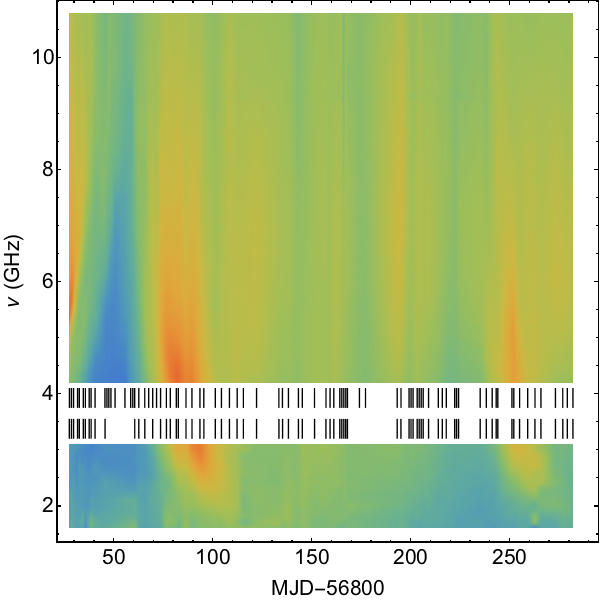}\hspace{3mm}
\put(5,20){\includegraphics[height=63mm]{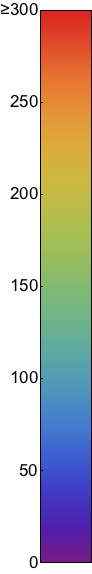}}
\caption{Observed flux density of PKS 1939-315 (in $\mathrm{mJy}$, as per the colour bar) as a function of Modified Julian Date and frequency.
{\mods 
The horizontal, white strip is the gap between the $4\,\mathrm{cm}$ and $16\,\mathrm{cm}$ bands in our observing configuration.
The upper (lower) set of tickmarks in this strip denote the epochs of observation in the $4\,\mathrm{cm}$ ($16\,\mathrm{cm}$) band.
Between observing epochs, we use linear interpolation to determine appropriate flux densities.
The time axis in this figure is related to the apparent source position, $\beta$, via the adopted kinematic model.
}}
\label{figure:characteristics}
\end{figure}

We applied the methods presented in section~\ref{subsection:reconstruction} to our Australia Telescope Compact Array data on the ESE towards \source. This event was discovered via a novel method utilising the wide bandwidth of the Compact Array Broadband Backend (CABB, \citealt{cabbpaper}). The method of discovery and the follow-up data are described by \cite{bigpaper}. The data consist of flux density measurements taken in 16\,cm  ($1.6-3.1\,\mathrm{GHz}$), and 4\,cm ($4.2-10.8\,\mathrm{GHz}$), bands with $4\,\mathrm{MHz}$-wide spectral channels.\footnote{The original data covered bandwidths slightly larger than the  values quoted here. We trimmed the band edges, as the extrema are difficult to calibrate accurately.}  The source was observed every few days after the ESE was identified in June 2014 until March 2015. Gaps in this coverage lasting up to a month are present, particularly in the 16\,cm band. {\mods The data are presented in Figure~\ref{figure:characteristics}.}

{\mods
We tried the different methods described in section 3.3 when reconstructing the characteristic system of the lens mapping $\theta(\beta,\lambda)$, for both extremely anisotropic and axisymmetric cases, but only the method of Algorithm~\ref{algorithm:trials} proved useful in practice. We tested for missing high- or de-magnification events in gaps in temporal coverage by breaking the data into separate pieces, but we found no conclusive evidence for any. The test instead revealed a small but consistent lag of change in $\theta$ compared to the predictions of $\mu$, as if the latter was biased low. We interpret the bias as due to the presence of a steady component that is not subject to lensing, either because of its large size, or because it is substantially offset from the lensed component of the source. To counter the bias we introduced into our models a parameter, $f$, to describe the unlensed fraction of the source flux. To keep the spectral model simple we adopted the same unlensed fraction at all frequencies; the optimal values found were $f=0.04\pm0.01$ for the extremely anisotropic and $f=0.05\pm0.01$ for axisymmetric geometries. We thus employ a total of three model parameters to describe the source.
}

In the vicinity of caustics, a point-source model yields arbitrarily large fluxes and the root-mean-square flux difference is overwhelmingly influenced by these regions. Root-mean-square flux difference is therefore not a good statistic for assessing the goodness-of-fit of a model. A better choice of demerit is to use the root-mean-square difference in the inverse-flux:
\begin{equation}\label{fod}
\mathrm{demerit}=\left\{\frac{\sum\left[F_0^{-1}(\lambda_j)\mu^{-1}(\beta_i, \lambda_j)-F^{-1}(t_i, \lambda_j)\right]^2}{\#\mathrm{data~points}-\#\mathrm{model~parameters}}\right\}^{1/2},
\end{equation}
which we used in our optimisation. We also quote the r.m.s. of the flux density residuals, to permit comparison with measurement noise levels. {\mods Although only a single number, $\theta_1$, is freely adjusted during the optimisation in Algorithm~\ref{algorithm:trials}, we include in our parameter count the total number of characteristics in the model, which is 86. We do so because for each characteristic the gradient of the lens mapping is chosen so as to match, in an average sense, the data close to that characteristic. This also allows us to compare models $\alpha_0(\theta)\sim\{(\alpha,\theta)_i\}$ with varying resolution and regardless of how they were obtained.}
 
The refractive index of a plasma increases strongly with wavelength, as per equation~(\ref{agkscale}), leading to multiple refracted images, and to diffractive scattering at sufficiently low frequencies. Moreover, the angular size of compact radio quasars tends to increase with wavelength. These considerations suggest that the methods discussed above are best suited to short radio wavelengths. We therefore focused our modelling effort exclusively on the 4\,cm band, with the 16\,cm band data being used as an independent check on the model. We present our model predictions in the 16\,cm band, below, for illustration only; {\momods we compare them to the data  and discuss the discrepancies in Section~\ref{section:discussion}.}

\begin{figure*}
\centering
\includegraphics[height=70mm]{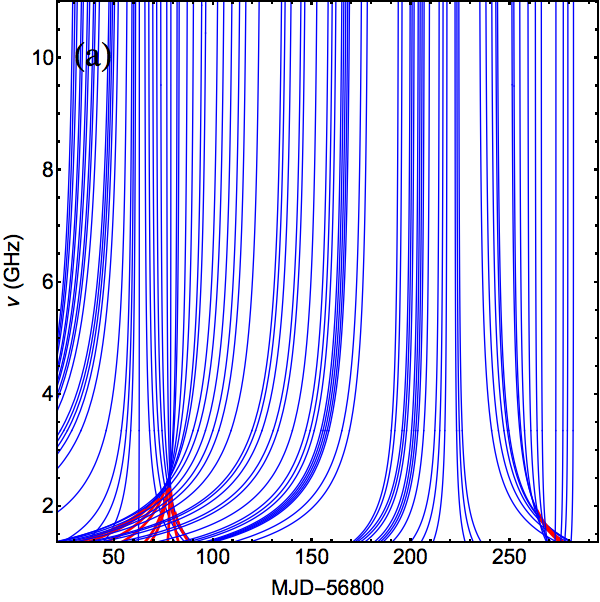}\ \ \ \includegraphics[height=70mm]{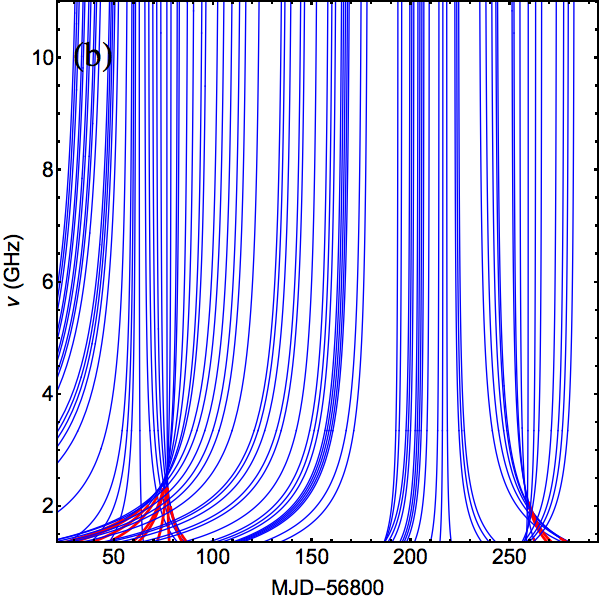}

\includegraphics[height=70mm]{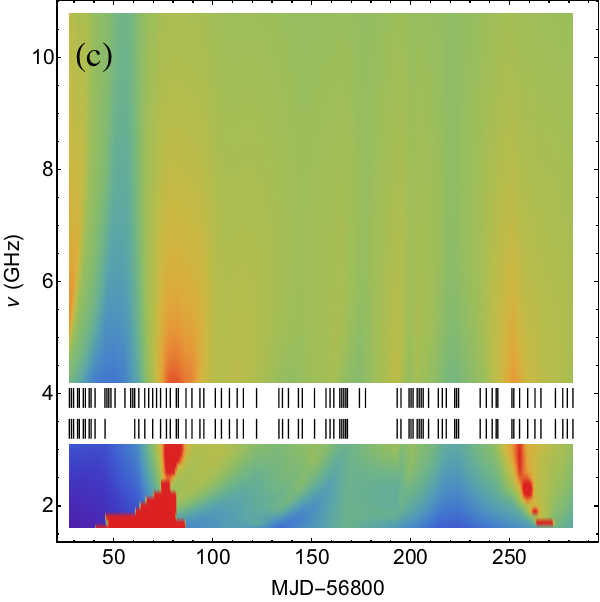}\ \ \ \includegraphics[height=70mm]{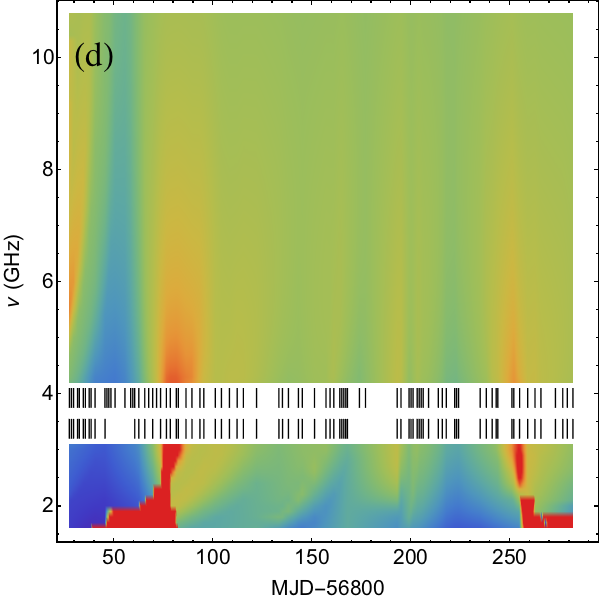}
\caption{{\mods Characteristic curves (upper panels: a, b), and dynamic spectra (lower panels: c, d), for the optimal extremely anisotropic model (left panels: a, c), and the optimal axisymmetric model (right panels: b, d).  The characteristic curves are plotted in blue for positive parity and red for negative; the red-blue boundary exhibits the caustic structure of these models. In order to permit a direct visual comparison with figure 1, the model dynamic spectra are constructed with identical time and frequency sampling as our data, and use the same colour scale as figure 1. At a frequency of $8.4\,$GHz, the locations of the characteristic curves coincide with the $4\,$cm band observing epochs.}}
\label{figure:models}
\end{figure*}

{\mods
The characteristic lines used in modelling cannot coincide with the data at all frequencies. To estimate flux densities on characteristic points that fall in between the data sampling locations (and vice versa), we used linear interpolation in $(\beta, \lambda)$ coordinates\footnote{{\mods Specifically, we interpolated in $\lambda$ first -- at $\beta=\mathrm{const}$ for data, or along the characteristics for models -- and then interpolated in $\beta$.\label{footnote:interpolation}}}. When computing the average of equation~(\ref{1dmuscale}), which depends non-linearly on the observable flux densities, $F$, we weighted the estimates at each wavelength in proportion to the (squared) derivative of $\partial_\beta\theta_0$ with respect to $F$, so as to avoid biasing the average by a few estimates with a low denominator. This presumes uniform uncertainty in $F$ along the spectrum, which should be a good approximation for the thermal noise in our data. 

For computational reasons, we averaged the data over $\sim25$ spectral channels, yielding a regular spectral grid sampled at 100~MHz intervals. We chose all the characteristic pivot points at $8.4\,\mathrm{GHz}$, and placed them at $\beta$ positions corresponding to the 86 epochs available in the 4\,cm band between June 2014 and March 2015. This choice minimises the influence of interpolation, because bending angles are small at high radio-frequencies --- as can be seen from the characteristic curves plotted in Figure~\ref{figure:models}.
}

\subsection{Modelling results}
\label{subsection:modellingresults}

The results of fitting for both the extremely anisotropic and axisymmetric geometries are quite similar, both qualitatively and in terms of the demerit figures, and we cannot decide between these geometries. Table~\ref{table:models} summarises the performance of the resulting models and Figure~\ref{figure:models} shows the predicted dynamic spectra. Both models produce a reasonable fit to the data in the 4\,cm band. Nevertheless, the uncertainties on our data are significantly lower than the r.m.s. flux residual between model and data, 
\footnote{With channels of 100 MHz, the thermal noise is only $\sim0.6\,$mJy and the uncertainties in the data are dominated by residual calibration errors and source confusion, which we expect to be of order 1\% (i.e. $\sim2\,$mJy).}
and therefore there is considerable scope for improvement in the models.  At lower frequencies the residuals are much worse and exceed the average flux density of the source itself -- a result of the very high magnifications in the vicinity of the model caustics.

\begin{figure}
\centering
\includegraphics[height=80mm]{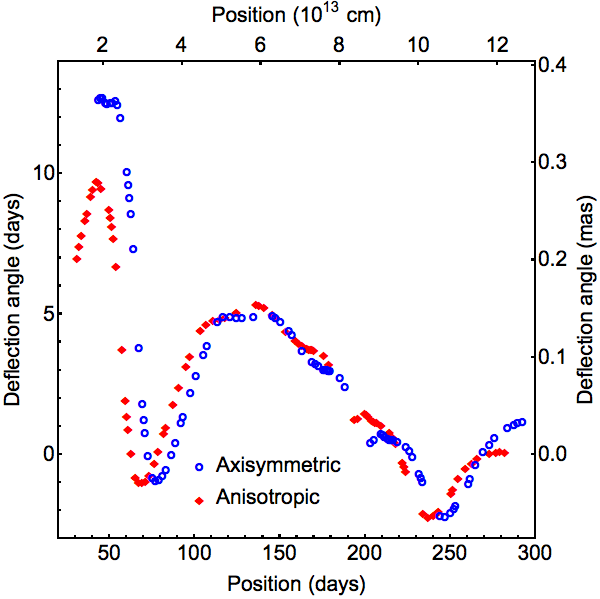}
\caption{{\momods Deflection angles $\alpha_0$ of the two optimal models shown for the reference frequency $\nu_0=5.9\,\mathrm{GHz}$. The horizontal axis is position in the lens plane, $D_d\theta$, with $\theta=\beta+\alpha$. The natural modelling units are shown on the left and bottom axes,} whereas the right and top axes assume the effective velocity and distance of  $v_f=50\,\mathrm{km}\,\mathrm{s}^{-1}$, $D_d=1\,\mathrm{kpc}$ for conversion to physical units. The latter scale as $v_f$ and $v_f/D_d$ on the horizontal and vertical axes,  respectively. The origin coincides with the position of the source at the reference epoch MJD56800 for the anisotropic model, and the symmetry centre for the axisymmetric model.}
\label{figure:deflectioncurves}
\end{figure}

\begin{table}
\caption{Parameters of the optimal models}
\label{table:models}
\centering
\begin{tabular}{rll}
 & Anisotropic & Axisymmetric \cr
\hline
\#model parameters & 86+3 & 86+3+2 \cr
demerit ($\mathrm{Jy}^{-1}$), 4\,cm& $0.27$ & $0.27$ \cr
r.m.s. residual flux (mJy), 4\,cm  & $7.1$ & $7.2$ \cr
demerit ($\mathrm{Jy}^{-1}$), 16\,cm & $9.7$ & $6.0$ \cr
r.m.s. residual flux (mJy), 16\,cm & $1000$ & $720$\cr 
\hline
\end{tabular}
\end{table}

\begin{figure}
\centering
\includegraphics[height=80mm]{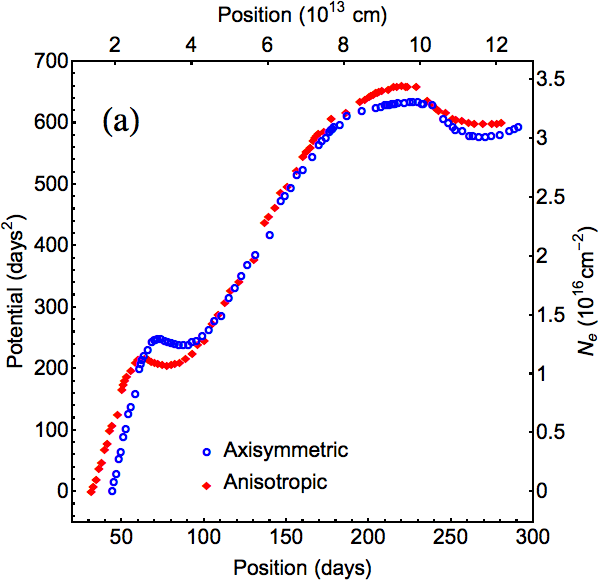}

\includegraphics[height=80mm]{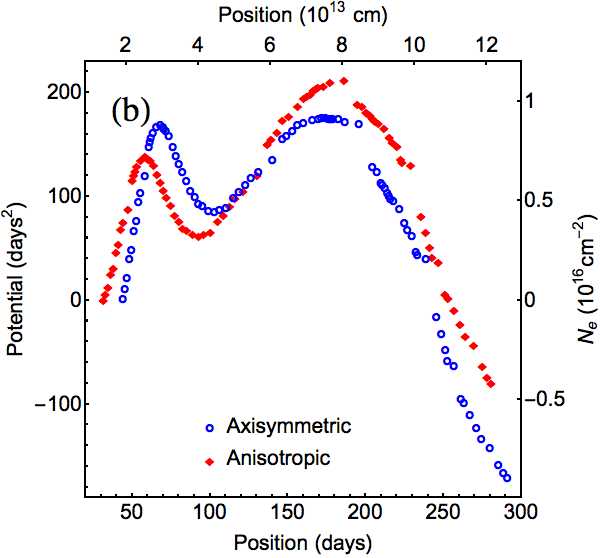}

\caption{Panel (a) shows the potential function, {\momods $\psi\propto N_e$, obtained by integrating the deflection curves in Figure~\ref{figure:deflectioncurves}}. Panel (b) shows the same curves with the mean 
gradient subtracted, {\momods in order to show more clearly the lumps and bumps which are present.  The horizontal axis is position in the lens plane, as in Figure~\ref{figure:deflectioncurves}. The potential is presented in the natural modelling units on the left-hand vertical axis. The right-hand  axis shows the physical units corresponding to $v_f=50\,\mathrm{km}\,\mathrm{s}^{-1}$ and $D_d=1\,\mathrm{kpc}$ and scales as $v_f^2/D_d$.}}
\label{figure:Nes}
\end{figure}

Figure~\ref{figure:deflectioncurves} presents the deflection curves of the two characteristic sets. The curves are presented both in the ``modelling units'' of days ({\it i.e.}, with $v_f=1$ and $D_d=1$), and for illustration we have also converted to physical units by assuming specific values for the effective velocity and distance: $v_f=50\,\mathrm{km}\,\mathrm{s}^{-1}$ and $D_d=1\,\mathrm{kpc}$, respectively. Integrating the deflection curves via equations~(\ref{Psidef}, \ref{psidef}) we obtain the potential function, $\psi$, which is proportional to the electron column density~(\ref{psiNe}).

Figure~\ref{figure:Nes} presents the inferred potential on top, and the bottom panel shows the same estimates with the mean gradient of $\psi$ removed {\momods(so as to reveal the lumps and bumps which are superposed on the trend).} The gradient does not affect magnification and would be impossible to measure at a single frequency. It does affect the alignment between light-curve features at different wavelengths, making it detectable with spectral data. The mean gradients that we infer through averaging deflection angles in Figure~\ref{figure:deflectioncurves} are
\begin{eqnarray}\label{meangradients}
\left\langle\nabla N_e\right\rangle\approx\left(\begin{array}{l}5.2\cr5.9\end{array}\right) 10^{16}\,\mathrm{cm}^{-2}\,\mathrm{yr}^{-1} \left(\frac{v_f}{50\,\mathrm{km}\,\mathrm{s}^{-1}}\right)^2\left(\frac{1\,\mathrm{kpc}}{D_d}\right)\nonumber\\
\approx\left(\begin{array}{l}330\cr380\end{array}\right)\,\mathrm{cm}^{-3} \left(\frac{v_f}{50\,\mathrm{km}\,\mathrm{s}^{-1}}\right)\left(\frac{1\,\mathrm{kpc}}{D_d}\right),\hspace{1.4cm}
\end{eqnarray}
where the upper and lower figures are for the extremely anisotropic and axisymmetric geometries, respectively. 

Our method is {\momods completely} insensitive to the density zero point {\momods so readers are free to add any constant vertical offset they like to the points in figure 4. On the other hand the curvature and gradient of the density directly affect the light curves and their alignment across the frequency dimension, so those quantities are fixed by the modelling at every plotted point in figure 4. 

{\momomods The uncertainties in mean-gradient and mean-curvature may be of interest. By varying the parameter $\theta_1$ we obtain models which have a different mean gradient from our optimum model, but which necessarily have a greater demerit. Because the uncertainties on our data are primarily systematic ($\sim2\,{\rm mJy}$), we can only give rough estimates of the uncertainty associated with model parameters; we adopt uncertainty intervals defined by the mean-square residual being no larger than its minimum value plus  $4\,{\rm mJy^2}$. With this criterion we find that the uncertainty on the mean-gradient in the lens is $\pm40$\%\ around the value given in equation (\ref{meangradients}).}

{\momomods Even a small change in the mean curvature away from our best lens model would, by itself, greatly increase the residuals of the fit, because of the change in the time-averaged spectrum of the lensed source.  However, the inferred mean curvature of the lens is, to some extent, degenerate with the shape of the intrinsic source spectrum, and the intrinsic spectrum is not precisely known. We therefore modified our source spectrum model in such a way as to precisely mimic the effect of magnification by a plasma lens, and then determined the mean convergence of the resulting lens model. As expected, the lens model tries to compensate for the change in the spectral model in such a way that the lensed source spectrum is largely unchanged. By varying the degree to which the intrinsic spectrum is modified, and evaluating the mean-square residuals of the resulting lens models we were thus able to estimate the uncertainty in the mean-convergence of the lens. We determined that the data permit an additive constant convergence of up to $\pm0.007$. In other words, the zero-point in figure~\ref{figure:convergence} is uncertain by only $\pm0.007$. The sensitivity of our data to a constant offset in the lens curvature stems, in part, from the large region of the lens-plane which we have sampled: a range in $\theta$ of order 250 days. Over this interval a constant curvature of only $0.01$ changes the gradient by about 5 days, which is roughly twice as large as the mean gradient in our best models --- see figures~\ref{figure:deflectioncurves} and~\ref{figure:Nes}.}

We emphasise that these constraints are strictly local and do not depend on the behaviour of the column-density at large distances from the line-of-sight --- there is no analogue of the ``external shear'' which is encountered in gravitational lensing. Nor is there an analogue of gravitational lensing's ``mass-sheet degeneracy'' --- that degeneracy is broken by the strong frequency-dependence of the plasma refractive index. For example: the large negative curvature in $\psi$ (hence $N_e$) around day 60 in Figure~\ref{figure:Nes} is responsible for the period of strong demagnification around MJD 56850 in Figure~\ref{figure:characteristics}.}

\section{Discussion}
\label{section:discussion}

Even with an ideal inversion algorithm, the dynamic spectrum alone cannot provide a complete characterisation of a plasma lens. Such an inversion returns $\psi(\theta)$, and the scaling factors required to obtain plasma column-density as a function of position (see section~\ref{section:lensing}) must be determined by other means. In common with previous authors we can, of course, simply assume values for the effective transverse velocity and lens distance, in order to fix those conversion factors. And, naturally, if we choose values similar to those used by previous authors then we obtain similar properties for the lenses themselves, because in all cases the basic observational requirement is to obtain magnification variations $\sim1$ for a few months. Thus if we adopt $v_f\sim50\,\mathrm{km}\,\mathrm{s}^{-1}$ and $D_d\sim1\,\mathrm{kpc}$, then we infer lensing by a structure of size $\sim1\,\mathrm{AU}$ with {\mods electron column-density gradient $\sim3\times10^2\,\mathrm{cm}^{-3}$. For an axisymmetric lens, which presumably arises from a structure that has spherical symmetry, the column-density gradient is expected to be comparable to the volume density of electrons in the lens, implying tiny lenses which are highly over-pressured. For highly anistropic lenses it is not obvious what the appropriate three-dimensional structure is, so the relationship between $n_e$ and $N_e^\prime$ remains unspecified. Many authors have argued in favour of sheet-like structures seen edge-on, as a means of moderating the required electron pressure \citep{rbc1987,goldreichsridhar2006,penking2012, penlevin2014}.}

In addition to ESEs, two other radio-wave scattering phenomena appear to require a large population of tiny structures incorporating ionised gas with large gradients in $N_e$. These phenomena are the Intra-Day Variability (IDV) of some compact radio quasars (e.g. \citealt{KCJ1997, dennettthorpedebruyn2000, bignalletal2003, masiv2}), and pulsar parabolic arcs \citep{stinebringetal2001, cordesetal2006}. These two phenomena can be plausibly attributed to a single type of scattering structure, differing only in the nature of the background radio source \citep{tbw2013}. There is good evidence that these structures are often very highly anisotropic in their scattering (e.g. \citealt{walkeretal2004, cordesetal2006, walkerdebruynbignall2009, brisken2010}), which suggests that strong magnetic fields may be present. In the modelling of \source\ reported in this paper we were unable to distinguish between axisymmetric and highly-anistropic lenses. However, it is notable that our best-fit axisymmetric model demanded a very small impact parameter for the event (consistent with zero), so that the relative orientation of the effective velocity vector and the plasma density contours is essentially constant during our monitoring interval. Put another way: when we ask for an axisymmetric model, the data prefer us to choose one which looks very much like an extremely anisotropic model. This deduction, made from fitting the dynamic spectrum alone, is consistent with the astrometric shifts of \source, measured with the Very Long Baseline Array \citep{bigpaper}, which are all consistent with refraction in the same direction on the sky. We note, however, that the interpretation of the observed astrometric shifts is unclear at present (see \citealt{bigpaper}).

One further aspect of our data is worth drawing attention to. In the dynamic spectrum of \source, there appear to be flux variations at all times. Thus, although there is an interval of outstandingly strong variability in our dynamic spectrum of \source\ -- i.e. the two months either side of MJD56850, which triggered our follow-up observations \citep{bigpaper} -- that interval should perhaps not be thought of as an isolated event. Rather, it might be better thought of as a large example of the ongoing fluctuations which are currently manifest in this particular souce.

Although the foregoing points are suggestive, not conclusive, they are all consistent with the Extreme Scattering ``Event'' in \source\ being closely related to the episodic phenomenon of Intra-Day Variability in radio quasars, and thus also to the pulsar parabolic arcs. It has previously been noted that in pulsar parabolic arcs one sometimes sees deflection angles consistent with a lens that is strong enough to explain the large magnification fluctuations observed in ESEs \citep{hilletal2005}.

A valuable, independent check on the quality of our modelling is provided by our ATCA 16\,cm band data on \source, which were not utilised in constructing our model plasma lens profiles. {\momods The lensing effects of a plasma increase with wavelength, and caustics are expected to form at some point.}  Comparison of Figure~\ref{figure:models} with Figure~\ref{figure:characteristics} immediately reveals the deficiencies of our models, particularly in the poor correspondence between the locations of model caustics and local peaks in observed flux density. {\momods Nor do the peaks in the data appear strong and sharp, as expected for caustics.} These qualitative deficiencies are quantitatively reinforced by the figures of demerit and r.m.s. flux residuals given in table~\ref{table:models}. The poor performance of our models at low-frequencies is not surprising, for two reasons. First, because the optical-depth to synchrotron self-absorption decreases with frequency, radio quasars are most compact at high-frequencies, and thus our point-source approximation is a poorer approximation at low-frequencies. Indeed, VLBA observations reveal \citep{bigpaper} that the resolved jet component contributes only a few per cent of the total flux density at $8.4\,\mathrm{GHz}$ but at $1.4\,\mathrm{GHz}$ and $2.3\,\mathrm{GHz}$ the resolved component dominates the compact core. Although the foregoing point is not necessarily responsible for errors in the location of the caustics, the figures of demerit and r.m.s. flux residuals are both strongly influenced by the presence of caustics when the moderating influence of source-size vanishes. Second, our treatment is based on geometric optics and thus excludes the effects of diffraction by small-scale  column-density structures in the lens. {\momods Diffraction is also expected to be more important at low frequencies because the phase profile of the lens scales in proportion to the wavelength (\ref{PhiNe}), and because the Fresnel scale, which separates refractive and diffractive regimes, increases with wavelength.} Unfortunately, non-zero source size and diffraction both appear to be incompatible with the core ideas of this paper -- i.e. characteristics, and a well-defined scaling of magnification with wavelength -- so the exclusion of these physical effects is a fundamental limitation of our method. On the other hand forward modelling that includes these phenomena is feasible, which suggests that parameterisation of the lens and source structures, followed by optimisation, should be a profitable avenue to explore.

Our method of determining the lens profile provides built-in consistency checks~(\ref{llinv}, \ref{ooinv}) that can be applied to the reconstructed characteristic sets, by estimating the optical scalars from the fit to equation~(\ref{muscale}) along each characteristic. We performed these checks, fitting for the coefficients of $\lambda^2$ and $\lambda^4$ in equation~(\ref{muscale}), and the models largely satisfy their consistency relations. But the uncertainty and level of degeneracy between the two coefficients preclude us from drawing any further conclusions. The uncertainties are in fact so large that the optical scalars inferred for the axisymmetric model easily satisfy the consistency check for the anistropic model.

\begin{figure}
\centering
\includegraphics[height=80mm]{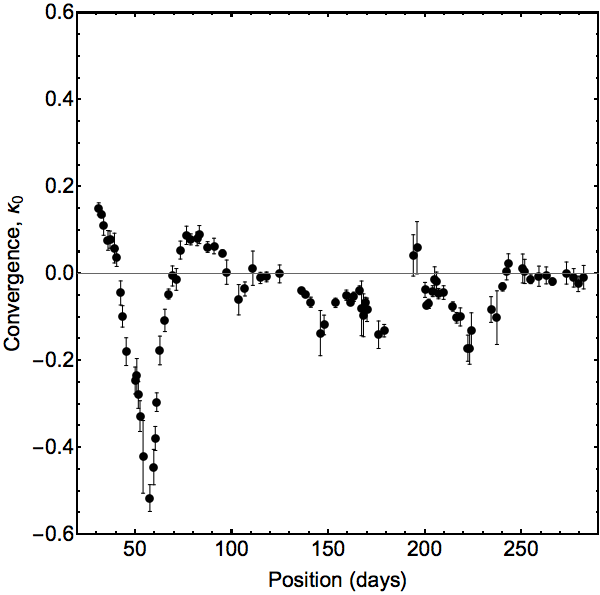}
\caption{{\mods Beam convergence (at the reference frequency $\nu_0=5.9\,\mathrm{GHz}$) of our optimal extremely anisotropic model computed as the average of equation~(\ref{llinv}) along each characteristic. {\momods The horizontal axis is the position in the lens plane, as in Figure~\ref{figure:deflectioncurves}}. The  convergence is clearly negative in the course of the strong demagnification event. The error bars show the r.m.s. variation of the individual $\kappa_0$ estimates along the characteristic.}}
\label{figure:convergence}
\end{figure}

{\mods
The results of our modelling are nevertheless very definite regarding the nature of the lens responsible for the period of strongest demagnification. As shown in Figure~\ref{figure:convergence}, it was caused by a diverging lens, with $\kappa_0<0$.} 
\cite{penking2012} argued that the intervals of strong demagnification observed during ESEs are  naturally explained by a converging lens that is over-focused. Our lens inversion is predicated on the assumption that only a single image is present at high frequencies, and does not directly address the circumstance of an over-focused lens (which is necessarily in the caustic regime). However, towards the upper end of the frequency range of our data (Figure~\ref{figure:characteristics}) the fractional variations in flux are both small and decreasing, making it difficult to sustain the idea that this lens is over-focused.

Attempting to extend our approach, which identifies characteristics by minimising the deviation from equation~(\ref{muscale}), to the case of a general two-dimensional lens incurs the penalty of a broadened interpretation of both the solution and the characteristic lines. It can be shown that the general reconstruction problem in 2D does not have a unique solution, and therefore one is obliged to look for particular classes of solution or their representatives. Likewise, a one-dimensional line~(\ref{lenscharacter}) in the three-dimensional model space $(\boldsymbol\theta, \lambda)$ would not generally \emph{lie on} the surface defined by the two-dimensional data $(t, \lambda)$ but instead only \emph{intersects} this surface. However, the lines on this surface that minimise the deviation from equation~(\ref{muscale}) do exist, and by construction they project to intervals $\boldsymbol\theta(\lambda)$ with little variation in the optical scalars. This makes it possible, in principle, to constrain the physical properties of the screen in both transverse coordinates.

\section{Conclusions}
\label{section:conclusions}
The advent of new, high quality data on plasma lensing events motivates the development of novel techniques which exploit the information in those data to constrain the lens properties. We have presented one such method, which takes a dynamic spectrum as input and yields the plasma column-density profile as output. In so far as possible, our approach avoids the proliferation of model freedoms: we use geometric optics, the thin-screen approximation, a point-source model, and a lens with reduced dimensionality. These restrictions are all of questionable validity and one or more may need to be relaxed in future. Nevertheless, restricting the solution space in this way permits a straightforward and rapid exploration of the lens properties, providing immediate insights into lens structure and a firm foundation for more sophisticated modelling. 

Applying our method to ATCA data on the ESE in \source\ we recovered profiles of the lens responsible for this event in the case of either axisymmetric or extremely anisotropic lens geometries. The two profiles share many features, including a large mean gradient and a column-density peak associated with the region of strongest lensing. Both our models provide a good visual match to the high frequency data (4.2-10.8~GHz), which were used as input, but neither did a good job of predicting the lower-frequency data (1.6-3.1~GHz), which were used only as a test of the models. There is, therefore, considerable scope for improvement in the inversion procedure.

\acknowledgments
The Australia Telescope Compact Array is part of the Australia Telescope National Facility, which is funded by the Commonwealth of Australia for operation as a National Facility, and managed by CSIRO. We especially thank the scheduler, Phil Edwards, for making the regular ATCA observations possible. We have benefitted greatly from the guidance of Ron Ekers on a variety of issues relating to this work.


\end{document}